\documentclass[10pt]{article}
\usepackage[cp1251]{inputenc}
\usepackage[english]{babel}
\begin{document}
\begin{center}
\textbf{MODELLING COSMIC ACCELERATION IN MODIFIED YANG - MILLS THEORY}\\
\vspace{2,5mm} \emph{\bf{ V.~K.~Shchigolev} \footnote{E-mail:
vkshch@yahoo.com}}
\end{center}
\begin{center}
Department of Theoretical Physics, Faculty of Physics and\\
Engineering, Ulyanovsk State University, Ulyanovsk 432000, Russia
\end{center}
\vspace{3,5mm}

\abstract{\noindent We investigate the possibility that the
modified Yang-Mills theories can produce an accelerated cosmic
expansion. We take into account some specific non-trivial solution
of the modified Yang-Mills equation  obtained by the author
earlier, which allows us to build several modifications of
accelerated cosmic expansion.}

\vspace{3,5mm}


\vspace{2,5mm}


\section {\large Introduction}
\qquad Presently it is accepted by the scientific community that
the Universe is experiencing an accelerated expansion. This is
supported by many cosmological observations, such as SNe Ia
\cite{C1}, WMAP \cite{C2}, SDSS \cite{C3} and X-ray \cite{C4}.
These observations suggest that the Universe is dominated by the
Dark Energy (DE), which provides the dynamical mechanism for the
accelerated expansion of the Universe. Moreover, they suggest that
the DE equation-of-state (EoS) parameter, $\displaystyle \gamma
=\frac{p}{\rho}$\,  might have crossed the phantom divide $\gamma
= -1$ from above in the near past. In order to explain this
phenomena, one can either consider theories of modified gravity,
or field models of DE.  The simplest candidate of DE is a tiny
positive time-independent cosmological constant, for which
$\gamma=-1$. However, it is difficult to understand why the
cosmological constant is about 120 orders of magnitude smaller
than its natural expectation (the Planck energy density). This is
the so-called cosmological constant problem. Another puzzle of DE
is the cosmological coincidence problem: why are we living in an
epoch in which the dark energy density and the dust matter energy
are comparable? As a possible solution to these problems various
dynamical models of DE have been proposed, such as quintessence
\cite{C5}.  So far, a large class of scalar-field DE models have
been studied, including tachyon \cite{C6}, ghost condensate
\cite{C7} and quintom \cite{C8}, and so forth. In addition, other
proposals on DE include interacting DE models \cite{C9},
braneworld models \cite{C10}, and holographic DE modeles
\cite{C11}, etc. The quintom scenario of DE is designed to
understand the nature of DE with gamma across -1. The quintom
models of DE differ from the quintessence, phantom and k-essence
and so on in the determination of the cosmological evolution.

Another class of DE models is based on the conjecture that a
vector field can be the origin of DE \cite{C12},\cite{C13}. The YM
field can be a kind of candidate for such a vector field
\cite{C14},\cite{C15}. At the same time, it is well known that a
pure YM field (with its EoS $\gamma = 1/3$) can not provide
accelerated expansion of the Universe, for which $\gamma < 1/3$ is
required. This is a direct consequence of conformal symmetry of
the Lagrangian for a massless YM field. Any violations of
conformal symmetry (e.g., as a result of quantum corrections
\cite{C16} or of non-minimal coupling to gravity \cite{C17}) give
a good chance for involving YM fields in reconstruction of DE. The
alternative method for YM fields to be involved in DE problem is
consideration of some sort of modified YM theory
\cite{C18}-\cite{C20}. In this paper, we turn our attention to the
issue of the YM fields as a source of DE in the frame of a
modified YM theory. We aimed at deriving the necessary conditions
for the possibility of those models to explain the accelerated
expansion of the Universe.  We also derive the corresponding
equations for the scale factor evolution, and briefly discuss
several examples for the DE models of this kind.

\section {\large Equations of model dynamics}

\qquad Let  us consider the following action \cite{C20}:
\begin{equation}
\label {1} S = -\int d^4x
\sqrt{-g}\Bigl\{\frac{R-2\Lambda}{2\kappa}+\Phi(F^a_{ik}F^{aik})\Bigr\},
\end{equation}
where $F^a_{ik}=\partial_iW^a_k - \partial_kW^a_i +
f^{abc}W^b_iW^c_k$, and $\Phi$ is a continuously differentiable
function. Variation of (\ref{1}) with respect to $g^{ik}$ yields
the Einstein equation
\begin{equation}
\label {2} G_{ik}\equiv R_{ik}-\frac{1}{2}g_{ik}R = 2\kappa
\Bigr(\frac{1}{2}g_{ik}\Phi- 2 \Phi'\, F^a_{ij}F^{aj}_k\Bigl),
\end{equation}
where $\Phi' \equiv d \Phi (I)/d I$, and $I = F^a_{ik}F^{aik}$ is
the invariant of the Yang-Mills fields. As it follows from the
action (\ref{1}), the equation of motion for the field potential
$W^a_i$ turns into
\begin{equation}
\label {3}\partial_i \Bigl( \sqrt{-g}\,\Phi'
F^{aik}\Bigr)+\sqrt{-g}\,\Phi'\, \epsilon^{abc}W^b_i F^{cik}=0.
\end{equation}

We assume that the Universe is described by a
Friedmann-Robertson-Walker (FRW) geometry:
\begin{equation}
\label {4} d s^2 = N(t)^2 d t^2- a^2 (t)(d r^2+\xi^2 (r)d \Omega
^2),
\end{equation}
where $\xi (r)=\sin r,r,\sinh r$ for the sign of space curvature
$k=+1,0,-1$, consequently. To study a FRW solution of the
equations (\ref{2}),(\ref{3}), we can directly insert this metrics
into action (\ref{1}). Whereupon we obtain the following
effective Lagrangian density (per unit solid angle):
\begin{equation}
\label {5} L_{eff}=\frac{3}{8\pi G}\Bigl( -\frac{a \dot
a^2}{N}+kaN-\frac{\Lambda a^3}{3}N\Bigr)\xi^2 - \Phi(I)a^3 N
\xi^2.
\end{equation}
At the same time, the generalized Wu-Yang ansatz for the $SO_3$ YM
fields can be written as \cite{C21}
$$
W^a_0 = x^a \frac{W(r,t)}{er},
$$
$$W^a_{\mu} =\varepsilon_{\mu
ab}x^b\frac{K(r,t)-1}{er^2}+\Bigl(\delta^a_{\mu}-\frac{x^ax_{\mu}}{r^2}
\Bigr)\frac{S(r,t)}{er}.
$$
We can make the following substitution into this ansatz
\cite{C22}:
$$
K(r,t)= P(r)\cos \alpha(t),~S(r,t)= P(r)\sin
\alpha(t),~W(r,t)= \dot \alpha(t).
$$
As a result, we have the following formulae for the YM strength
tensor components:
\begin{eqnarray}
{\bf F}_{01}={\bf F}_{02}={\bf F}_{03}=0,~~~~~~ {\bf
F}_{12}=e^{-1} P'(r)\Bigl({\bf m }\,\cos\alpha + {\bf
l}\,\sin\alpha\Bigr),\nonumber\\
{\bf F}_{13}=e^{-1} P'(r)\sin\theta\Bigl({\bf m }\,\sin\alpha-{\bf
l}\,\cos\alpha\Bigr),~~~ {\bf
F}_{23}=e^{-1}\sin\theta\Bigl(P^2(r)-1\Bigr){\bf n}, \label {6}
\end{eqnarray}
where \,\,$ {\bf n}= (\sin\theta \cos \phi, \sin \theta \sin \phi,
\cos \theta),~~ {\bf l}= (\cos\theta \cos \phi, \cos \theta \sin
\phi, \cos \theta)$ and ${\bf m}= (-\sin \phi, \cos \phi, 0) $ are
the orthonormalized isoframe vectors, and the prime means a
derivative with respect to $r$. As noted in \cite{C22}, the YM
field (\ref{6}) has only magnetic components. It is easy to find
from (\ref{4}) and (\ref{6}) that the YM field invariant $I =
F^a_{ik}F^{aik}$ becomes as follows:
\begin{equation}
\label {7} I = \frac{2}{e^2 a^4 \xi^2}\Bigl[ 2 P'^2+\frac{(P^2
-1)^2}{\xi^2} \Bigr].
\end{equation}
Varying the effective Lagrangian density (\ref{5}) over $P(r)$,
and taking into account (\ref{7}) we obtain the following
Euler-Lagrange equation instead of YM equation (\ref{3}):
\begin{equation}
\label {8} \Bigl\{ P''-\frac{(P^2 -1)P}{\xi^2} \Bigr\}\Phi' +P'
\Phi''\frac{2}{e^2 a^4 }\Bigl\{ \frac{1}{\xi^2}\Bigl[ 2 P'^2
+\frac{(P^2 -1)^2}{\xi^2}\Bigr] \Bigr\}'= 0,
\end{equation}
where $\Phi'\equiv d \Phi(I)/d I,\, \Phi''\equiv d^2 \Phi(I)/d I^2
$.

The nontrivial solution for the YM equation obtained in
\cite{C22}, $P(r)= \xi'(r)=\cos r, \cosh r$ for $k=+1,-1$,
consequently, satisfies  equation (\ref{8}). Indeed, this solution
turns both additive terms in the left-hand-side of (\ref{8}) to
zero. As it follows from (\ref{7}), the valuable feature of this
solution is that the YM invariant built on it depends only on
time:
\begin{equation}
\label {9} I = I(t) = \frac{6}{e^2 a^4(t)}.
\end{equation}
The Hamiltonian constraint for (\ref{5}) is
\begin{equation}
\label {10} \Bigl(\frac{\dot a}{a}\Bigr)^2 + \frac{k}{a^2}=
\frac{8\pi G}{3}\Phi (I)+ \frac{\Lambda}{3},
\end{equation}
where $I$ should be replaced with its value (\ref{9}). By
variation of (\ref{5}) over $a(t)$ with the subsequent choice of
the gauge $N=1$, one can obtain the following Friedmann equation:
\begin{equation}
\label {11} 2\frac{\ddot a}{a}+\Bigl( \frac{\dot a}{a}\Bigr)^2 +
\frac{k}{a^2}= 8\pi G \Phi(I)-\frac{64 \pi G}{e^2 a^4}\Phi'(I)+
\Lambda.
\end{equation}

\section {\large Accelerated expansion}

\qquad First we have to note that instead of equation (\ref {11})
one frequently uses the following equation:
\begin{equation}
\label {12} \frac{\ddot a}{a}= \frac{8\pi G}{3}\Bigl[\Phi(I) -2 I
\Phi'(I)\Big]+
 \frac{\Lambda}{3},
\end{equation}
which can be obtained by combining two equations (\ref {10}),
(\ref {11}), and with the help of (\ref {9}). This equation is just
the differential consequence of equation (\ref {10}).
Nevertheless, it is convenient for investigation of the
accelerated expansion.

Comparing equations (\ref{10}) and (\ref{12}) with the similar
ones of the standard FRW cosmology of a perfect fluid, we can find
the following expressions for the effective energy density and
pressure of the YM field and the cosmological constant:
\begin{equation}
\label {13} \rho = \Phi(I)+\frac{\Lambda}{8\pi G},~~~p = -\Phi(I)
+ \frac{4}{3}\,I \, \Phi'(I)-\frac{\Lambda}{8\pi G},
\end{equation}
where the last terms are just the energy density and pressure
associated with cosmological constant $\Lambda$ with EoS:
$\rho_{\Lambda}=-p_{\Lambda}=\displaystyle \frac{\Lambda}{8 \pi
G}$. Therefor, the EoS for the YM field and cosmological constant
in our model is
\begin{equation}
\label {14} \gamma = -1 +
\frac{4}{3}\,\frac{I\,\,\Phi'(I)}{\Phi(I)+\displaystyle
\frac{\Lambda}{8\pi G}}.
\end{equation}
As it follows from equation (\ref{12}) or, equivalently, from the
inequality $\gamma < -1/3$ in (\ref{14}), the accelerating regime
is possible if
\begin{equation}
\label {15} \Phi(I) -2 I \Phi'(I)+
 \frac{\Lambda}{8 \pi G}>0.
\end{equation}

Now we are going to consider some interesting examples concerning
the standard and modified YM theories.

{\bf a)} For the standard YM theory
$\Phi(I)=\displaystyle\frac{1}{16\pi}I$, that is
$\Phi(I)=\displaystyle \frac{3}{8\pi e^2 a^4(t)}$. Plugging this
$\Phi(I)$ into (\ref{14}) we have
\begin{equation}
\label {16} \gamma = -1 + \frac{4}{3}\,\Bigl(1+\displaystyle
\frac{\Lambda e^2}{3 G}a^4(t)\Bigr)^{-1}.
\end{equation}
This  gives the EoS $\gamma = 1/3$ in the case of vanishing
$\Lambda$ as it must be for the pure radiation.  Besides, the
accelerating condition (\ref{15}) turns into
\begin{equation}
\label {17} a(t)>a_c=(3 G/\Lambda e^2)^{1/4}
\end{equation}
when $\Lambda \ne 0$. In this case, EoS (\ref{16}) goes from
$1/3$ to $-1$ during the evolution of the scale factor $a(t)$, and
becomes less then $-1/3$ as (\ref{17}) is satisfied. Plugging
$\Phi(I)=\displaystyle \frac{3}{8\pi e^2 a^4(t)}$ into (\ref{10})
we have the following equation for the scale factor:
$$
\dot a ^2 + k= \frac{G}{e^2 a^2}+ \frac{\Lambda}{3}a^2,
$$
which can be easily  integrated. It should be noted that the
similar equation was discussed  earlier in \cite{C17}.

{\bf b)} Let us now suppose the phenomenological power-law
dependence of $\Phi(I)$ on $I$: $\Phi(I)= A\,I^n$, where $A,n$ are
some nonzero constants. In this case, one can rewrite inequality
(\ref{15}) as
\begin{equation}
\label {18} (1-2n)\,A\,I^n + \frac{\Lambda}{8 \pi G}>0.
\end{equation}
The latter means $n<1/2$ in the case of vanishing cosmological
constant: $\Lambda = 0$. At the same time, according to (\ref{14})
the EoS becomes $\gamma= -1+\displaystyle \frac{4}{3}n =
constant$. For $n<1/2$, $\gamma < -\displaystyle \frac{1}{3}$ that
is this model experiences eternal accelerated expansion.

Let us revert to the case of non-zero $\Lambda$. As it follows
from (\ref{9}) and (\ref{14}),
\begin{equation}
\label {19} \gamma_n = -1+\frac{4}{3}\,n
\Bigl(1+\displaystyle\frac{\Lambda e^{2n}}{8\pi G A
6^n}a^{4n}(t)\Bigr)^{-1}.
\end{equation}
With the help of (\ref{18}), it is easy to show that $\gamma_n < -1/3$
in this case too. Of course, the particular case $n=1,\,
A=1/16\pi$ leads to (\ref{16}). Now plugging
$\Phi(I)=\displaystyle A\frac{6^n}{e^{2n} a^{4n}(t)}$ into
(\ref{10}) we have the following equation for the scale factor:
\begin{equation}
\label {20} \dot a ^2 + k= B a^{2(1-2n)}+ \frac{\Lambda}{3}a^2,
\end{equation}
where $B=8\pi G A 6^n/3e^{2n}$. This equation can be
integrated for several $n$ in an explicit form.

{\bf c)} Now we consider the case of a widely discussed
non-Abelian Born-Infeld (BI) Lagrangian (see, e.g., \cite{C17} and
bibliography therein):
$$
L_{NBI}=\frac{\beta ^2}{4 \pi}\Bigl(\sqrt{1+\frac{F_{ik}^a
F^{aik}}{\beta^2}-\frac{(\tilde{F}_{ik}^a
F^{aik})^2}{16\beta^4}}\,-1\Bigr),
$$
where $\beta$ is the critical BI field strength,
$\tilde{F}_{ik}^a$ is a dual YM strength tensor. From (\ref{6}),
we can find that for our solution the second invariant of YM field
$\tilde{\bf F}_{ik}{\bf F}^{ik}=0$.  Hence, we can identify
$\Phi(I)$ with
$$
\Phi(I)=\frac{1}{16 \pi \alpha}\Bigl(\sqrt{1+2\alpha I}\,-1\Bigr),
$$
where $\alpha = 1/2\beta^2$. Under our solution for YM field, this
model goes from $\gamma = -1/3$ to $\gamma = -1$ according to the
following equation:
\begin{equation}
\label {21} \gamma_{\,_{BI}} = -1 + \frac{8\alpha}{e^2}\Bigl(
a^4+\frac{12\alpha}{e^2}\Bigr)^{-1/2}\Bigl[\sqrt{a^4+\frac{12\alpha}{e^2}}+
\Bigl(\alpha\frac{\Lambda }{G}-1\Bigr)a^2 \Bigr]^{-1}.
\end{equation}
Simultaneously, the scale factor is driven by the equation
\begin{equation}
\label {22} \dot a ^2 + k =
\frac{G}{3\alpha}\Bigl(\sqrt{a^4+\frac{12\alpha}{e^2}}-a^2 \Bigr)+
\frac{\Lambda}{3}a^2.
\end{equation}
It should be noted that the
same equation was obtained earlier in non-linear BI theory on the
brane \cite{C17}.

{\bf d)} At last, let us consider  the effective YM field cosmic
model based on the effective Lagrangian up to 1-loop order
\cite{C16}, \cite{C17}, \cite{C23}. In our notation, this
Lagrangian can be written as follows:
\begin{equation}
\label {23} \Phi(I)=-\frac{b}{4}\, I \ln \Bigl|\frac{I}{2\kappa^2}
\Bigr|,
\end{equation}
where $\kappa$ is the renormalization scale of dimension of
squared mass, $b=11/12\pi^2$ is the Callan-Symanzik coefficient
for the generic gauge group considered here. From (\ref{21}), we
can find that $ \Phi' = -(b/4)\,( \ln |I/2\kappa^2|+1)$. Due to
this formula together with (\ref{9}) and (\ref{14}), we obtain
the following EoS:
\begin{equation}
\label {24}
\gamma_{\,_{YMC}}=\frac{1}{3}-\frac{4}{3}\,\frac{1+\displaystyle\frac{\Lambda
e^2}{12\pi G b}a^4(t)}{\displaystyle\ln
\Bigl|\frac{(e\kappa)^2}{3}a^4(t) \Bigr|+\frac{\Lambda e^2}{12\pi G
b}a^4(t)},
\end{equation}
which displays more complicated behavior on time then the ones
considered above. Indeed, it starts from $\gamma_{\,_{YMC}}=1/3$
at $a=0$ but approaches $\gamma_{\,_{YMC}}=-1$ through the break
of its continuity at $a(t)=a_{cr}$, which can be found from the
following transcendent equation:
$$
a^4_{cr}(t)=\frac{3}{(e\kappa)^2}\exp{\Bigl\{-\frac{\Lambda
e^2}{12\pi G b}a^4_{cr}(t)\Bigr\}}.
$$
Nevertheless, if the model starts its expansion at $a_0> a_{cr}$
(non-singular model), then no problem of that kind occurs. Taking
into account (\ref{9}), (\ref{10}) and (\ref{23}) we can find that
the scale factor of the model is driven by the following equation:
\begin{equation}
\label {25} (\dot a)^2 + k= \frac{4\pi G b}{e^2 a^2}\ln
\Bigl|\frac{(e\kappa)^2 a^4(t)}{3} \Bigr|+ \frac{\Lambda}{3}a^2.
\end{equation}
In the case of vanishing $\Lambda$, the EoS of this model follows
from (\ref{24}) as
$$
\gamma^0_{\,_{YMC}}= \frac{1}{3}-\frac{4}{3}\Bigl(\ln
\Bigl|\frac{(e\kappa)^2 a^4(t)}{3} \Bigr| \Bigl)^{-1}.
$$
The critical value $a(t)$ becomes
$a^0_{cr}=(3/e^2\kappa^2)^{1/4}\ne 0$.

Finally, we have to note that EoS (\ref{19}), (\ref{21}),
(\ref{24}) and corresponding equations (\ref{20}), (\ref{22}),
(\ref{25}) derived in the sub-sections {\bf b)}, {\bf c)} and {\bf
d)} consequently can be investigated in more details. That
could be done analytically or, in any case, numerically. We do not
make it our aim in this short communication.

\section{\large Conclusion}

\qquad In summary, the standard and some modified YM theories in FRW
cosmology are studied in this paper. The specific non-trivial
solution of the modified YM equation (\ref{8}) proposed  by the
author earlier allows us to build several modifications of
accelerated cosmic expansion. All of them possess EoS $\gamma \sim
-1$ at late time, so the cosmic coincidence problem can be avoided
in those models. Besides, we have derived the equations for the
cosmic scale factor in all those models.  In our opinion,  more
significant result of our study is that we can now to consider the
wide range of modified YM theories in cosmology. For such a purpose,
the equations (\ref{9},\ref{10}), (\ref{13},\ref{14}) and inequality
(\ref{15}) have to be employed. Further details and consequences of
the modified YM models considered here are in progress.

\qquad

\end{document}